# A universal description of III-V/Si epitaxial growth processes

I. Lucci[1], S. Charbonnier[2], L. Pedesseau[1], M. Vallet[3], L. Cerutti[4], J.-B. Rodriguez[4], E. Tournié[4], R. Bernard[1], A. Létoublon[1], N. Bertru[1], A. Le Corre[1], S. Rennesson[5], F. Semond[5], G. Patriarche[6], L. Largeau[6], P. Turban[1], A. Ponchet[3] and C. Cornet[1,*]

[1]Univ Rennes, INSA Rennes, CNRS, Institut FOTON – UMR 6082, F-35000 Rennes, France
[2]Univ Rennes, CNRS, IPR (Institut de Physique de Rennes) - UMR 6251, F-35000 Rennes, France
[3]CEMES-CNRS, Université de Toulouse, UPS, 29 rue Jeanne Marvig, BP 94347 Toulouse Cedex 04, France
[4]IES, Univ. Montpellier, CNRS, Montpellier, France
[5]Université Côte d'Azur, CRHEA-CNRS, Rue Bernard Grégory, F-06560 Valbonne, France
[6]Centre de nanosciences et de Nanotechnologies, site de Marcoussis, CNRS, Université Paris Sud, Université Paris Saclay, route de Nozay, 91460 Marcoussis, France

Here, we experimentally and theoretically clarify III-V/Si crystal growth processes. Atomically-resolved microscopy shows that mono-domain 3D islands are observed at the early stages of AlSb, AlN and GaP epitaxy on Si, independently of misfit. It is also shown that complete III-V/Si wetting cannot be achieved in most III-V/Si systems. Surface/interface contributions to the free energy variations are found to be prominent over strain relief processes. We finally propose a general and unified description of III-V/Si growth processes, including the description of antiphase boundaries formation.

Integrating monolithically III-V semiconductors on group IV ones is often considered as the ultimate step for co-integration of photonics with electronics, such as lasers, passive devices, or multijunctions solar cells [1,2]. The main issues of polar on non-polar epitaxy to overcome were soon identified in the 80's [3], [4]. But since the interplay between 3D growth mode, strain relaxation, antiphase domains and other defects was never clarified, researchers preferentially developed defects filtering strategies using thick III-V buffers grown on silicon [5]. Reaching higher photonic integration level now requires a deep understanding of the processes involved at the early stages of III-V/Si heterogeneous epitaxy.

Summarizing the large literature on the subject is hopeless, but we would like to emphasize on three major physical concepts about III-V/Si growth that are usually presented as implicit underlying statement and that are in close relationship with the present work.

First, the origin of AntiPhase Domains (APDs) formation is commonly attributed to either Si single steps or uncomplete group III or group V initial coverage of the Si surface. This general picture, described in details by Kroemer [3], is today considered as the main motivation for using misoriented Si substrates, in order to promote bi-step formation, and theoretically hamper the formation of antiphase boundaries.

Second, the origin of the commonly observed 3D islanding during III-V/Si growth was frequently ascribed to strain relaxation processes, for instance in the case of GaAs on Si [4,6], since most III-V semiconductors are lattice mismatched to the silicon. It was also noticed that for mismatched semiconductors significant densities of dislocations are generated well before island coalescence. However, 3D islanding was also already reported in quasi lattice-matched systems such as GaP/Si [7].

Finally, III-V/Si interface atomic arrangement was theoretically addressed on the basis of Density Functional Theory (DFT) calculations. This was for instance discussed in GaAs/Si [8] or more recently in GaP/Si [9–11]. Highlights were given on the fact that abrupt III-Si or V-Si interfaces are not always the most stable configurations, depending on the group-III/group-V chemical potentials. Indeed, some charge-compensated interdiffused interfaces following the electron counting model criteria [12] were found to be remarkably stable. [9,10,13,14]

In this letter, we aim to clarify the main III-V/Si crystal growth processes. From atomically-resolved microscopy analysis, the morphologies of mono-domain III-V (AlSb, GaP or AlN) islands at the Si (001 or 111) surfaces are first established. On the basis of absolute surface/interface energies calculated by ab initio (DFT) calculations on GaP/Si, the wetting properties are determined over the full range of phosphorus chemical potential. The respective contributions of surface/interface and stress relief to free energy variation during the III-V/Si epitaxy are then compared. We finally describe the main steps of the III-V/Si heteroepitaxy and the formation of antiphase domains.

3D islanding is first investigated through three different III-V semiconductor materials because they allow to span the initial epitaxial stress from compressive (AlSb/Si) to tensile (AlN/Si) through near-zero (GaP/Si).

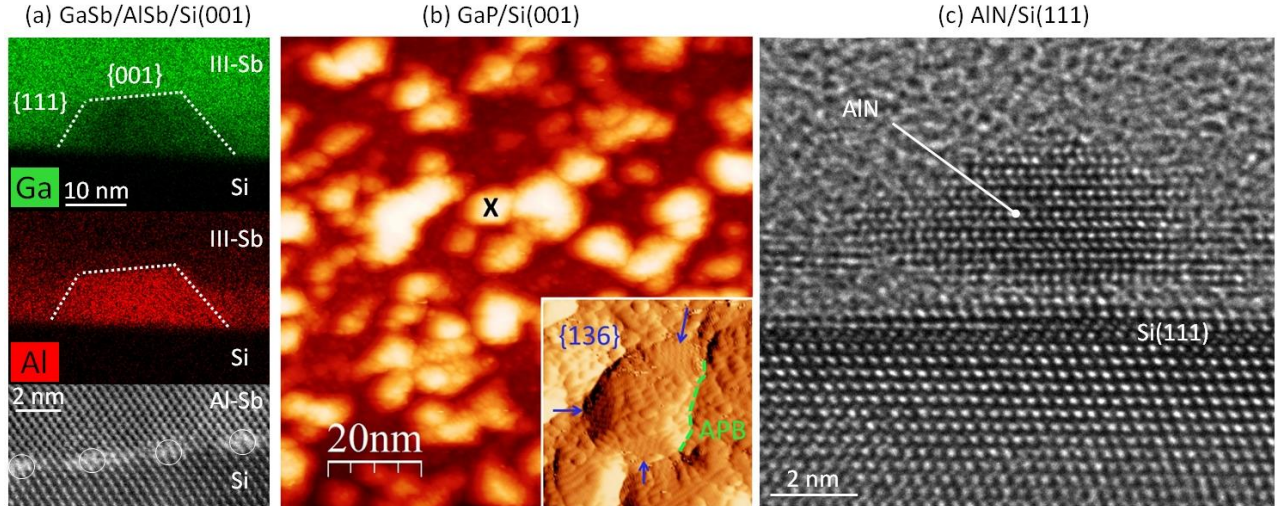

FIG. 1: 3D islanding in various III-V/Si materials systems. (a) Cross-sectional STEM-EDX image of GaSb/AlSb layers grown on Si (001) – 6°-off, showing the Ga and Al concentrations, and high resolution STEM imaging of the AlSb/Si interface, dislocations are surrounded. (b) plan-view STM imaging of a 3nm-thick GaP deposition on Si (001) – 6°-off. Inset shows the atomically-resolved morphology of the individual island marked with a black cross, with {136} facets and an antiphase boundary. (c) Cross-sectional high resolution TEM image of a 2nm-thick AlN deposition on Si(111).

In Fig. 1(a), the Scanning Transmission Electron Microscopy (STEM)-EDX images are given for AlSb/Si-6°-off islands (5 nm), buried in a GaSb matrix, with corresponding Ga- and Al- contrasts. High resolution TEM image of the interfacial misfit dislocations network is also given. Fig. 1(b) displays the scanning tunneling microscopy (STM) in-plane image of a 3nm-GaP/Si-6°-off deposition, a very early stage of growth, as compared to previous studies [15,16]. Inset shows the atomically-resolved typical morphology of one individual island at the surface, where {136} facets can unambiguously be identified [17], together with a trench being an antiphase boundary emergence. Fig. 1(c) shows the cross-sectional high resolution TEM image of a 2nm-AlN/Si(111) deposition. Experimental details on growth and microscopy are given in the supplemental materials [18].

From these experiments, some important conclusions can already be given. Firstly, in the various experiments performed on the three materials systems, 3D islands were always observed, and the presence of a wetting layer was not clearly or systematically evidenced (see Fig. 1(a) and (c) for instance), which confirms the partial wetting of the III-V on Si, i.e. the Volmer-Weber growth mode, independently of the strain state. [6,7] We believe that this is a general behavior of III-V/Si heteroepitaxial systems even when alternated growth techniques are used [18,19]. We will strengthen this assumption later on. Finally, the III-V/Si Volmer-Weber growth mode does not *a priori* hamper the Si surface to be terminated with a single monoatomic layer of group-III, group-V or other element rising from the epitaxial reactor background. Impact of such passivating layer will be discussed later.

It is also remarkable that in both AlSb and AlN materials systems, the misfit is so large that the III-V material relaxes very rapidly. Even if the relaxation process is not similar in Sb-based and N-based materials, the complete strain relief is nearly achieved at only 1 nm of the interface. Fig. 1(a) also illustrates that the island size is much larger than typical distances between dislocations. It was already reported that dislocations appear well before islands coalescence [6], and we note that the observed islands are nearly perfectly facetted well after crystal plastic relaxation. This suggests that elastic relaxation of strain [20] is not contributing significantly to individual islands energy balance. Here we conclude that surface/interface energies play a crucial role in III-V/Si 3D islanding.

The last important conclusion that can be drawn from experiments, is the mono-domain character of the observed single islands. In Fig. 1(b), most of the individual grains have a homogeneous morphology. The largest homogeneous islands (without APDs) are likely the consequence of smaller islands coalescence. Neighboring smaller islands are also visible, with a clear separation between them that seems to hamper the coalescence (shown with the green dashed line in Fig. 1(b) inset). The atomic structure of one individual island shown in the inset of Fig. 1(b) evidences the mono-domain character of the island and the presence of {136} facets. Therefore, from cross-sectional TEM and plan-view STM experiments it is clear that individual III-V/Si islands remain mono-domain. This observation is in agreement with the work of Akahane *et al*. [21] where individual AlSb or GaSb islands on Si were observed. Anisotropy of individual islands was demonstrated along either the [110] or the [1-10] silicon crystallographic axis, demonstrating the mono-domain character of single islands, and the overall bi-domain distribution of the islands population. The size of islands presented in Fig. 1 (a) and (b) is also interesting. Both GaP and AlSb epilayers were grown on Si(001)-6°-off substrates, where atomic (bi-atomic) steps are separated in average by 1.29 (2.58) nm. Mono-domain islands are significantly larger (≈10 nm), which contradicts the usual correlation made between mono-atomic Si steps and APBs formation [3].

To complete the picture, we note that the average spacing between islands (10 nm) in Fig. 1(b) corresponds well to the APDs correlation length measured on thicker epilayers grown under the same conditions ([8-12] nm) [22]. Finally, impact of III-V islands coalescence on III-V/Si epilayers structural quality was highlighted [7,23].

In a first and general description, the III-V/Si wetting properties can be examined within the Young-Dupré spreading parameter $\Omega$ [24]:

$$\Omega = \gamma^S_{(Si)} - \gamma^S_{(III-V)} - \gamma^i_{(III-V/Si)} \quad (1)$$

Where $\gamma^S_{(III-V)}$ and $\gamma^S_{(Si)}$ are the surface energies of the most stable III-V facet that would be involved in the 2D growth on the substrate and of the silicon surface respectively, $\gamma^i_{(III-V/Si)}$ is the interface energy between the III-V semiconductor and the Si. A positive value of $\Omega$ corresponds to perfect wetting conditions, while a negative value corresponds to partial wetting, i.e. a Volmer-Weber growth, or perfect non-wetting conditions. However, the evaluation of $\Omega$ requires the accurate determination of surface and interface energies, which is done for GaP in this work.

To this aim, different absolute surfaces and interface energies of interest were computed via DFT calculations (see the supplemental materials [18]). The silicon surface energy, was already widely discussed [25–27]. Silicon surfaces with or without steps have been considered in this work, and we find that the presence of steps at the silicon surface (at least for a miscut below or equal to 6°) does not change significantly the silicon surface energy range ([87-93] meV/Å²). For GaP, the situation is different, as the surface energies depend on the reconstruction of the facet, on the chemical potential, and therefore on the growth conditions used (P-rich or Ga-rich). Calculations show that {136} surface energies of the GaP are in the same range than {001} ones, as already found for GaAs [28]. Finally, abrupt Ga-Si or P-Si (001) GaP/Si interfaces energies also depend on the chemical potential [9,10]. In a first approximation, we do not consider the charge-compensated interfaces, that may further stabilize the interface [10]. The results obtained are summarized in Table I.

The spreading parameter $\Omega$ is then plotted in Fig. 2(a) as a function of the phosphorus chemical potential variation $\Delta\mu_P = \mu_P - \mu_P^{P-bulk}$ ($\mu_P$ is the chemical potential of P atoms, and $\mu_P^{P-bulk}$ is the chemical potential of P atoms in black phosphorus, see details in [18]), where the right (left) side corresponds to P-rich (Ga-rich) limit conditions [9].

Table I. GaP and Si surfaces and interfaces energies computed by DFT

| Surface/ interface | details | reconstruction | Energy (meV/Å²) | |
|---|---|---|---|---|
| | | | P-rich | Ga-rich |
| Si(001) | flat | c(2×4) | 92.8 | |
| Si(001) | $D_B$-step | p(2×2) | 89.3 | |
| Si(001) | $S_B$-step | p(2×2) | 89.2 | |
| Si(001) | $S_A$-step | c(2×4) | 87.1 | |
| GaP(001) | Rich-P | (2×4) | 57.4 | 72.4 |
| GaP(001) | Rich-Ga | (2×4)-md | 82.8 | 52.9 |
| GaP(136) | Type-A | (1×1) | 52.9 | 62.7 |
| GaP(136) | Type-B | (1×1) | 66.8 | 57.1 |
| GaP-Si | Abrupt Ga-Si | (1×1) | 72.0 | 40.8 |
| GaP-Si | Abrupt P-Si | (1×1) | 29.7 | 60.9 |

The calculation is presented both for the P-Si and the Ga-Si abrupt interfaces, with a $D_B$-stepped Si surface. The most stable {001} surface reconstruction was always considered at a given value of the chemical potential, explaining the slope variation of $\Omega$. Whatever the chemical potential and the interface, $\Omega$ remains negative, indicating partial wetting conditions, even if in extreme P-rich conditions with a P-Si abrupt interface, the DFT calculation accuracy does not allow to conclude unambiguously on the sign of $\Omega$ in this very narrow window. Considering that most III-V semiconductors have the same surface energies orders of magnitudes, this conclusion (partial-wetting conditions) can be extended to most III-V semiconductors deposited on Si. In the following, the abrupt Ga-Si interface will be chosen for illustration.

In Fig. 2(b), the spreading parameter is plotted as a function of the substrate surface energy in P-rich and Ga-rich conditions. $\Omega$ increases with the substrate surface energy, as expected by definition. In the same plot are also reported typical surface energies ranges of some commonly used starting Si surfaces (passivated or not) already considered in the literature, such as Si(001), Si(111), $SiH_2$, SiAs, SiP or $SiO_2$ (e.g. [29] or [30]). Here, SiX stands for X-terminated Si surface. Impact of surface pretreatment or orientation on interface energy is not taken into account. We here conclude that any Si surface pretreatment or passivation will tend to stabilize the highly reactive nude Si surface, and thus favor partial wetting conditions, strongly reducing the hope to reach complete III-V/Si wetting conditions in real epitaxial chambers where the passivation can be intentional or not.

To complete the picture at the sub-monolayer scale, and evaluate the relative contributions of stress relaxation and surface/interface energies, we now compare two different situations: a strained 2D GaP island (with a 1 monolayer height, growing laterally) and a relaxed 3D truncated pyramidal GaP island in its Wulf–Kaishew equilibrium shape growing in an homothetic way on the silicon substrate, as depicted in Fig. 2 (c). The shape is in good agreement with the one inferred from STM data of Fig. 1(b) [18].

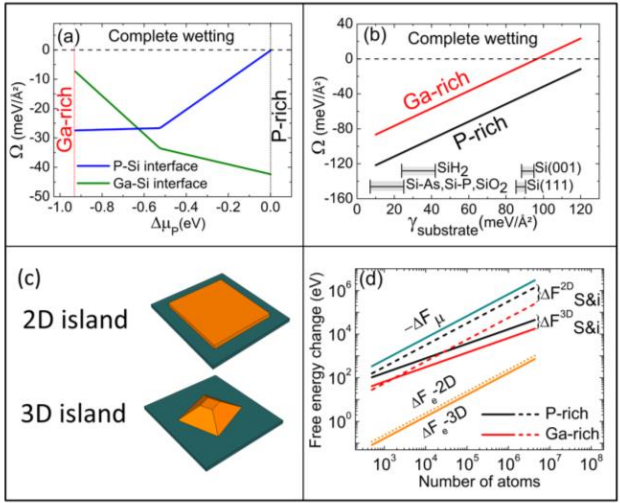

FIG. 2: (a) Spreading parameter vs the chemical potential variation for the deposition of GaP/Si, with P-Si and Ga-Si abrupt interfaces. (b) Spreading parameter vs substrate surface energy in P-rich conditions and Ga-rich conditions with Ga-Si interface. (c) Sketch of the 2D (strained) and 3D (elastically relaxed) GaP islands on Si. (d) The different contributions ($\Delta F_\mu$, $\Delta F_{S\&i}$, $\Delta F_e$) to the free energy variation for 3D and 2D GaP/Si islands with Ga-Si interface.

The total free energy variation during the GaP/Si growth is then calculated for the different 2D or 3D islands configurations by using [31]:

$$\Delta F_{TOT} = \Delta F_\mu + \Delta F_e + \Delta F_{S\&i} \qquad (2)$$

Details of the calculations are given in ref. [18]. The first term is the chemical work needed to form the bulk crystal from an infinite reservoir. The second term $\Delta F_e = R^* \Delta F_e\text{-}2D$ is related to the elastic energy stored, R being the relaxation energy factor, and $\Delta F_e$-2D the elastic energy of a biaxially strained 2D layer [31]. Here we take R=1 for the 2D GaP island growing on Si, and R=0.7 for the free elastic energy variation $\Delta F_e$-3D of the 3D GaP island [32]. The third term corresponds to the formation of surfaces and interfaces.

-$\Delta F_\mu$, $\Delta F_e$, $\Delta F_{S\&i}$ are plotted in Fig. 2(d) for both Ga-rich and P-rich conditions, and for the two types of islands, as a function of increasing number of atoms. The energy gain provided by the crystal formation $\Delta F_\mu$ is partly counter-balanced by both $\Delta F_e$ and $\Delta F_{S\&i}$, the elastic and surfaces/interfaces contributions. A first conclusion that can be drawn is that, whatever the phosphorus chemical potential, surface and interface energies have always a larger contribution to the energy variation than the elastic energy contribution. We also see that the contribution of the elastic energy is so weak that relaxation of strain has no impact on the island morphology which is thus mainly defined by surface/interface competition [18]. We finally evidence that, at small deposited number of atoms, 2D islands may be more stable than 3D ones. A precise description of this process would however require taking into account edge energies that is beyond the scope of this paper.

Importance of elasticity can be also discussed for other III-V semiconductors. For instance, the maximization of elastic energy in AlSb assuming a biaxial stress with R=0.005, leads to $\Delta F_e \approx 7.5 \cdot 10^2$ eV for $10^6$ atoms. This remains lower than typical surface/interface free energies variations. A significant contribution of misfit dislocations to interface energies is also expected in addition for mismatched systems. In the intermediate case of GaAs, where the relaxation occurs after some monolayers, elastic energy is expected to impact more seriously the island shape before the relaxation happens [31]. In any cases, after the plastic relaxation, surface and interface energy competition is clearly the most important contribution to the free energy variation, and has a prominent role for defining the shape of initial III-V/Si islands.

From these experimental and theoretical findings, it becomes clear that the physics of III-V/Si epitaxial growth is driven by the competition between III-V surface energies, Si surface energies and the III-V/Si interface energy. Main growth steps can be then derived and are represented in Fig. 3. Step (i): A thermal pretreatment of the Si surface possibly allows organizing Si steps (in mono or bi-atomic layers for (001) substrates), giving rise to a mono-domain or bi-domain distribution at the Si surface. A 35×35 nm² STM image of a Si(001)-6°-off surface is provided for a realistic illustration in Fig. 3(a). But the same process occurs on Si(111). Step (ii): The very reactive silicon surface is covered with a 2D complete or incomplete passivating layer (Fig. 3(b)). This can be accomplished intentionally with hydrogen for instance in chemical vapor deposition reactors, or unintentionally with growth chamber residual atmosphere exposure, group-V initial exposure such as Si-As, Si-N, Si-Sb or Si-P, or group-III initial exposure. This lowers the Si surface energy (see Fig. 2(b)), and promote partial wetting conditions. Step (iii): The nucleation starts and forms 2D or 3D small nuclei that can appear and disappear. This step is driven kinetically. The crystal polarity (we will use A and B to distinguish the two possible phases) of each nucleus is defined locally with respect to the silicon surface local orientation (Fig. 3(c)).

Step (iv): Stable 3D islands are formed and grow (Fig. 3(d)). Epitaxial relationship and (if necessary) dislocation network (including tilt, twist) are determined locally. Each island is mono-phase, because the energy cost to form an antiphase boundary is too large. Consequently, once an island is stable, its polarity is preserved during its subsequent growth by an adaptation of the charge-compensated interface structure, whatever the nature of the steps at the surface. The density of such stable islands directly defines the subsequent density/size of APDs. This density is fully determined by the kinetics of nucleation, [33] mainly imposed by group III atoms migration, i.e. growth temperature, nature of group III atoms used, V/III ratio, but also the vicinality used (numbers of steps at the surface), and the nature of the passivation layer at the Si surface.

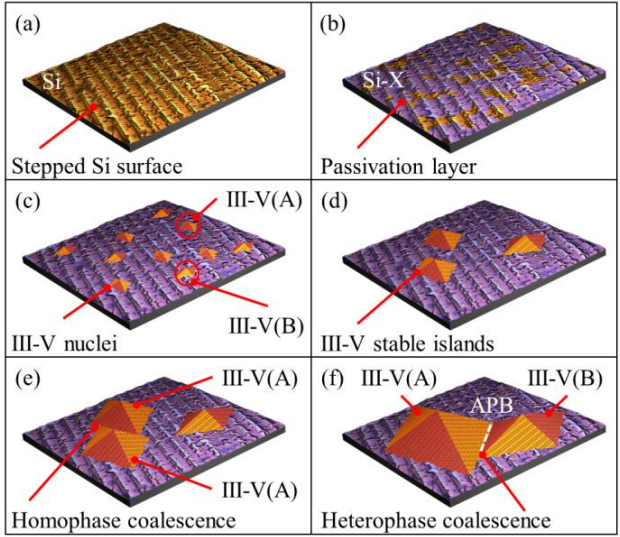

FIG. 3: Description of the proposed III-V/Si Growth steps, with (a) the 35×35 nm² STM image of a stepped starting Si surface. The Si surface is then covered (b) at least partially with a 2D passivation layer. Nucleation starts (c) with local epitaxial relationships and crystal polarity. Some stable islands then grow (d), independently of Si steps. If 2 islands of the same phase coalesce (e), they will form a larger island. If 2 islands having different phases coalesce (f), antiphase boundaries will appear.

The comparison between Al and Ga group III-atoms in ref. [21] perfectly illustrates this point. Kinetics also explains why APDs observed in the literature are usually larger (i.e. lower density) on nominal substrates than on vicinal ones, due to Ehrlich-Schwoebel barrier at step edges during diffusion processes.

Step (v): Islands cover a large part of the Si surface, and coalescence happens. If the two islands have the same phase, the homophase coalescence leads to the formation of a larger island (Fig. 3(e)). In this process, different tilt, twist and dislocation network structures within individual islands may impact the structural quality of the coalesced island. If the two islands have different phases, the heterophase coalescence necessarily leads, in addition to all the previous structural considerations, to the formation of an antiphase boundary (Fig. 3(b)). Generation of APDs in III-V/Si epilayers is thus governed by the respective area ratio of the different Si terraces orientations, and not to the monoatomic steps areal density as usually suggested [3].

Overall, we finally conclude that most of the structural defects usually formed during III-V/Si epitaxy (twist, tilt, imperfect dislocations networks or APDs) fundamentally originate from the partial wetting of III-V semiconductors on silicon, without a significant impact of elasticity. This generalized description of III-V/Si growth processes opens new routes to deeply co-integrate photonics and electronics.

The authors acknowledge Prof. P. Müller and Dr. I. Berbezier for fruitful discussions on the properties of silicon surfaces. Dr. J. Zhu is also acknowledged for giving advices on DFT description of polar surfaces. The authors would like to thank P. Vennéguès for the AlN/Si TEM image as well as IMRA Europe S.A. for the access to the JEOL 2100F microscope. This research was supported by the French National Research Agency ANTIPODE Project (Grant no. 14-CE26-0014-01), ORPHEUS Project (Grant no. ANR-17-CE24-0019-01) and Région Bretagne. The ab initio simulations have been performed on HPC resources of TGCC and CINES under the allocation 2017-[x20170906724] made by GENCI (Grand Equipement National de Calcul Intensif).

# A universal description of III-V/Si epitaxial growth processes
# Supplemental Materials
Revised 4/6/2018  19:01:00

I. Lucci[1], S. Charbonnier[2], L. Pedesseau[1], M. Vallet[3], L. Cerutti[4], J.-B. Rodriguez[4], E. Tournié[4], A. Létoublon[1], R. Bernard[1], N. Bertru[1], A. Le Corre[1], S. Rennesson[5], F. Semond[5], G. Patriarche[6], L. Largeau[6], P. Turban[1], A. Ponchet[3] and C. Cornet[1,*]

[1]UMR FOTON, CNRS, INSA-Rennes, F-35708 Rennes, France
[2]IPR, UMR 6251, CNRS-Université de Rennes I, Campus de Beaulieu 35042 Rennes Cedex, France
[3]CEMES-CNRS, Université de Toulouse, UPS, 29 rue Jeanne Marvig, BP 94347 Toulouse Cedex 04, France
[4]CNRS, IES, UMR 5214, F-34000 Montpellier, France
[5]Université Côte d'Azur, CRHEA-CNRS, Rue Bernard Grégory, F-06560 Valbonne, France
[6]CNRS, Laboratoire de Photonique et de Nanostructures, Route de Nozay, F-91460 Marcoussis, France

## GROWTH AND MICROSCOPY DETAILS

GaSb/AlSb/Si sample presented in Fig. 1(a):

The 6°-off (001) Si substrate was first prepared *ex situ* according to the procedure described in ref. [1] before being loaded into the MBE reactor. The substrate temperature was then ramped up to 800 °C at ~ 20 °C/min and then immediately cooled at the same rate down to 500 °C, without any intentional flux (all shutter cells being kept closed). MBE growth was initiated by simultaneous opening of Al and Sb shutters to grow 5 nm AlSb. Next, a thick GaSb layer was grown. The whole structure was grown at 500 °C, measured by a pyrometer, and the growth rates were 0.35 ML/s for AlSb and 0.65 ML/s for GaSb.

GaP/Si sample presented in Fig. 1(b):

GaP/Si sample presented in Fig. 1 (b) has been grown by Molecular Beam Epitaxy (MBE) on a HF-chemically prepared Si(001) substrate, with a 6° miscut toward the [110] direction [2]. The substrate has been heated at 800°C during 10 minutes to remove hydrogen at the surface, and a 3-nm thick GaP/Si deposition was performed at 350°C, with a subsequent 500°C short annealing and a cooling under phosphorus following the approach developed in previous studies [2–4]. An amorphous As capping layer was then deposited at cryogenic temperature, for the transfer of the sample to the Scanning Tunneling Microscopy (STM) experiment, already discussed in ref. [5].

AlN/Si sample presented in Fig. 1(c):

AlN/Si sample has been grown by MBE in a RIBER Compact 21S reactor. A cold-neck solid source is used for Al whereas ammonia is used as N source ($NH_3$-MBE). The nominal Si(111) substrate is HF-chemically prepared. After introduction in the growth chamber, the substrate is heated at 780°C to desorb hydrogen atoms, giving rise to a (7 x 7) surface reconstruction. In order to promote large and well defined terraces, the Si substrate was flashed at 1200°C (read by pyrometer). While cooling down, we observe serval orders of the (7 x 7) surface reconstruction, indicating that the Si surface is clean and well ordered. Then the AlN nucleation starts at 600°C, following the procedure described in ref. [6] , and the growth temperature is raised up to 1030°C. The AlN growth rate is of 100 nm/h. For the purpose of this study, a very thin AlN layer is grown without rotation, which results in a nominal thickness varying from 1.6 to 2.3-nm along the substrate diameter.

Scanning Tunneling Microscopy image of Fig. 1(b):

Scanning Tunneling Microscopy (STM) was performed at room-temperature in the constant current mode of operation. Tungsten electro-chemically etched tips were used. After MBE growth, an amorphous thick As capping layer was deposited on the GaP/Si(001) films at cryogenic temperature, allowing the transfer of the sample to the ultra-high vacuum STM chamber experiment, as already discussed in ref. [5]. Complete thermal desorption of the As protective layer was obtained at 500°C and allows STM observations of the GaP films. Raw STM images were simply corrected by subtraction of a basal plane. The (136) crystallographic planes of the GaP island facets were unambiguously identified by measuring the facet angle with respect to the basal plane. This was further confirmed by identification of the atomic arrangement of the (136) facets previously observed in ref. [5].

We also note here that alternated growth technique such as migration enhanced epitaxy (MEE) are sometimes used to promote the 2D planarity of the layers [7]. However STM measurements performed on GaP/Si suggest that alternated growth on the contrary leads to higher density of smaller islands, with earlier coalescence, which may be misinterpreted as a 2D layer in conventional resolution-limited atomic force microscopy.

In the following, it will be shown that {136} facets do not respect the electron counting model. Consequently, as already proposed in GaAs, the {136} facet reconstruction will be changed to the stable {2 5 11} one for the surface energy calculations. The very small difference of facets orientation remains fully compatible with the STM observations presented here. The full justification of this change will be explained in the DFT part.

Transmission Electron Microscopy image of Fig. 1(a):

The GaSb / Si sample (V2447) has been observed in cross-sectional view by Scanning Transmission Electron Microscopy on an aberration corrected microscope Titan Themis 200. The thin foil has been prepared by FIB following the <110> zone axis (the <110> direction parallel to the surface steps linked to the 6° misorientation). The FIB preparation has been following by a cleaning with argon milling at low voltage (1.5kV) during 9 minutes to remove the material redeposition (gallium and antimony) during the FIB process.

Transmission Electron Microscopy image of Fig. 1(c):

The AlN/Si sample for TEM observation is prepared using a conventional technique, involving mechanical thinning followed by ion-milling using $Ar^+$ at 0.5-5 keV, by pure mechanical wedge polishing or by focused ion beam. Cross-sectional view is observed in a JEOL 2100F microscope.

## DFT GENERAL DETAILS

All the calculations were performed within the Density Functional Theory [8,9] as implemented in SIESTA package [10,11] with a basis set of finite-range of numerical atomic orbitals. Calculations have been carried out with the generalized gradient approximation (GGA) functional in the Perdue-Burke-Ernzerhof (PBE) form [12] Troullier–Martins pseudopotentials [13], and a basis set of finite-range numerical pseudoatomic orbitals for the valence wave functions [14].

## SILICON

Silicon DFT details

To study by DFT the Si(001) surface energy, atomic relaxations have been done using a double-$\zeta$ polarized basis sets [15] with an energy shift of 50 meV and a real space mesh grid energy cutoff of 150 Rydberg. The geometries were optimized until the forces were smaller than $0.005 eV.\text{Å}^{-1}$. The electronic structure was converged using a (6 x6 x2) Monkhorst-Pack grid [16] in the case of the flat Si(001) surface and a (1x1x1) Monkhorst-Pack grid for the Si(001) stepped surface.

All the surfaces investigated have been modeled with the supercell approach which consists in a supercell made by a vacuum region and a periodic system (slab) in the (a,b) plane. The slab surfaces are orthogonal to the c axis. Actually, SIESTA is more suitable than plane-wave methods are to treat vacuum. Thus, a vacuum of 400Å has been chosen to avoid too much interaction between the periodic slabs. Each slab thickness is at least about 15Å. In the case of the Si(001) surface, the basis vectors are 15Å long. Instead, the $D_B$-step, $S_B$-step, $S_A$-step Si(001) surfaces consist into two terraces which extend over a rectangular surface whose long-side dimensions is 65.2Å, 38.6Å, 38.6Å respectively, while the short-side is 15.4Å in each one. The Si(001) surface is non-polar which means that the two slab surfaces are symmetric. That is why, in all the cases studied, almost 4Å on the top and on the bottom of the bulk were allowed to relax, while the atoms of the bulk have been frozen. The bulk lattice constant of 5.46Å was used in the silicon surface calculations.

Si(001) surface reconstruction

The Si(001) surface has been widely studied since decades [17–20]. Indeed, it is well known that the Si(001) (2x1) reconstruction is the most stable configuration which minimizes its surface energy [18,21,22]. In particular, the surface atoms arrange themselves in dimers aligned along the [110] or the [1-10] direction. As a consequence, two kinds of configurations form in each double- and single-step surface. They are called $D_A$, $D_B$ and $S_A$, $S_B$ respectively. The subscripts "$_A$" and "$_B$" are referred to the dimer orientation, perpendicular ("$_A$") and parallel ("$_B$") with respect to the step edge (as Chadi's convention [17,18] ). Furthermore, it has also been studied that to reduce the surface energy, the dimers buckling occurs by forming different kinds of reconstructions [21,23].

In this work, the flat and stepped surface energies have been determined. We considered for each case the most stable reconstruction already known from literature [21,23,24]. The surface reconstructions studied are depicted in Fig. S1. The atoms are represented in different colors as a function of the distance from the bulk. They consist in asymmetric (2x1) altering buckled dimers configurations which are more precisely: the c(4x2) in the case of the flat Si(001) [23,25] (Fig. S1a), the *rebonded* p(2x2) for both the $D_B$- and $S_B$-step Si(001) (Fig. S1b and c) and finally the c(4x2) in the case of the $S_A$-step Si(001) (Fig. S1d) [21]. The $D_A$-step surface has not been considered since it has been already proved in the literature that it is the most unstable among all the surfaces investigated [26]).

Si(001) surface energies calculations

The equation used to determine the surface energy is the following:

$$\gamma_{surf} = \frac{E_{slab} - N_{Si}\mu_{Si-bulk}}{2A} \quad (S1)$$

Where $E_{slab}$ is the total energy of the slab when its surfaces are reconstructed, $N_{Si}$ is the number of Si atoms in the slab, $\mu_{Si-bulk}$ is the silicon bulk chemical potential, $A$ is the surface unit area and the factor 2 is because of the surfaces symmetry.

**Table S1.** Si(001) surface energies computed by DFT

|  | Reconstruction | Surface Energy (meV/$A^2$) |
|---|---|---|
| Si(001) | c(4x2) | 92.8 |
| $S_B$-step Si(001) | p(2x2) | 89.1 |
| $S_A$-step Si(001) | c(4x2) | 87.1 |
| $D_B$-step Si(001) | p(2x2) | 89.3 |

The surface energy values for each reconstruction are reported in Table S1. In according to ref. [26], we found that the $S_A$-step c(4x2) Si(001) is the most stable reconstruction with a value of 87.1 meV/$A^2$. The $S_B$- and $D_B$-step *rebounded* p(2x2) Si(001) surfaces energies differ from each other of just 0.2 meV/$A^2$. The c(4x2) reconstruction of the flat Si(001) surface is the most unstable one, with a value of 92.8 meV/$A^2$. Nevertheless, one can clearly see that these energy values are very close to each other which means that the Si steps do not impact too much the surface energy values.

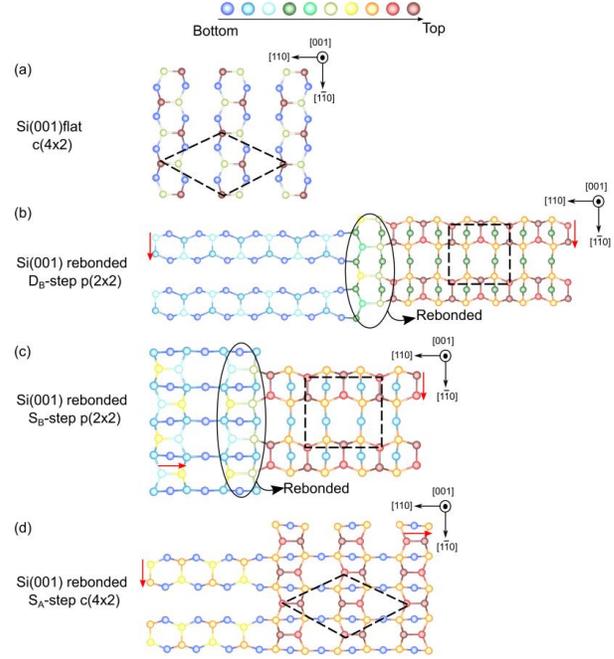

**Fig. S1:** Top view of the four silicon surfaces reconstructions investigated, which are: a) the c(4x2) of the Si (001) flat surface, b) and c) the *rebonded* p(2x2) of the $D_B$- and $S_B$-step Si(001) respectively and d) the c(4x2) in the case of the $S_A$-step Si(001). The dashed lines represent the reconstructions unit cells.

## GALLIUM PHOSPHIDE

GaP bulk, black phosphorus and the α-Ga phase DFT details

GaP, black phosphorus and the α-Ga phase have been modeled. A lattice constant of 5.57Å was used for GaP in the surface simulations. The black phosphorus and the α-Ga structures were used to estimate their own chemical potential to calculate the surface energy.

GaP (001) and (2 5 11) DFT details

In the case of the GaP(001) and GaP(2 5 11) surfaces, two different computational methods were used to determine their surface energies. The GaP(001) surface is non-polar, that is why an equation similar to Eq. S1 has been applied to this case. Instead, the GaP(2 5 11) is polar [27], therefore the bottom surface has been passivated with fictitious hydrogen atoms.

In terms of computational details, DFT calculations have been done using a basis set of finite-range numerical pseudoatomic orbitals for the valence wave functions [14]. $1s^{0.75}$, $1s^1$, $1s^{1.25}$, $3s^23p^3$, and $4s^23d^{10}4p^1$ were used as valence electrons for the fictitious H* with a net charge of 0.75e to compensate P, H, the fictitious H* with a net charge of 1.25e to compensate Ga, P, and Ga respectively. Structures relaxation and electronic structure calculations have been done using a double-ζ polarized basis sets [14] with an energy shift of 50 meV and a real space mesh grid energy cutoff of 150 Rydberg. The geometries were optimized until the forces were smaller than $0.005 eV.Å^{-1}$. The electronic structure was converged using 8x8x8, 8x2x6, 8x4x8, 2x2x1 and 3x2x1 Monkhorst-Pack grids [16] of the Brillouin zone for the GaP bulk, the black phosphorus, α-Ga phase and for GaP(001) and GaP(2 5 11) slabs respectively.

The two fictitious H*, H, Ga and P atoms have been built with ATOM code, the pseudopotential generation distributed as part of the SIESTA software package.

GaP(001) and GaP(2 5 11) slabs

The surface was modeled in a periodic slab geometry. The slab has been built to be periodic within the plan (a,b) and also to reveal the surface orthogonally to the c axis. Actually, SIESTA is more suitable than plane-wave methods are to treat vacuum. In this case, a vacuum of 450Å has been chosen for the same reasons as in the case of the Si surfaces. Each surface fulfills the electron counting model [28] as originally well-established for GaAs and ZnSe.

For the non-polar GaP(001) surfaces, the bottom and top surface have been treated identically with the same reconstruction which decreases the error on the determination of the surface energy. The thicknesses of the slab are about 17Å and 23Å respectively for the P-rich GaP(001)(2x4) surface (Fig. S2c,d) and for the Ga-rich GaP(001)md(2x4) surface (Fig. S2a,b). The subsurfaces of the top and bottom surfaces were allowed to relax about 6Å into their minimum energy configuration and all the others atoms were kept frozen in the bulk position.

Instead, for the polar GaP(136) surfaces, which have a thickness of about 20Å, we considered the reconstruction already studied for a similar case, which is the GaAs(2 5 11) [29–31]. This is due to the fact that the GaP(136) surface does not fulfill the ECM [28]. Therefore, it is necessary to add two P (Ga) atoms on the P-rich (Ga-rich) surface unit cell to respect it. As a consequence, three parallel slightly inclined P (Ga) dimers form in the surface unit cell, now lying on the (2 5 11) plane. The need to fulfill the ECM together with the fact that the (2 5 11) and (1 3 6) planes are very close (sustaining a leaning angle of ~2°) [30,32] make necessary working on the more stable GaP(2 5 11) surface rather than the (136). The two reconstructions are named P-rich GaP(2 5 11)A-(1x1) (Fig. S3a) and named Ga-rich GaP(2 5 11)B-(1x1) (Fig. S3b) (as shown in similar works in Ref. [29,30]). Also in this case, we used the letter A (B) referring to the P(Ga)-terminated surface.

The surfaces reconstructions were passivated by the fictitious H* with fractionally charged hydrogen 1.25e and 0.75e for Ga and P dangling bonds. Then, the subsurface opposite to the passivated surface of the slab was allowed to relax about 6Å into their minimum energy and all the others atoms were kept frozen in the bulk position except the fictitious H* atoms which were also allowed to relax.

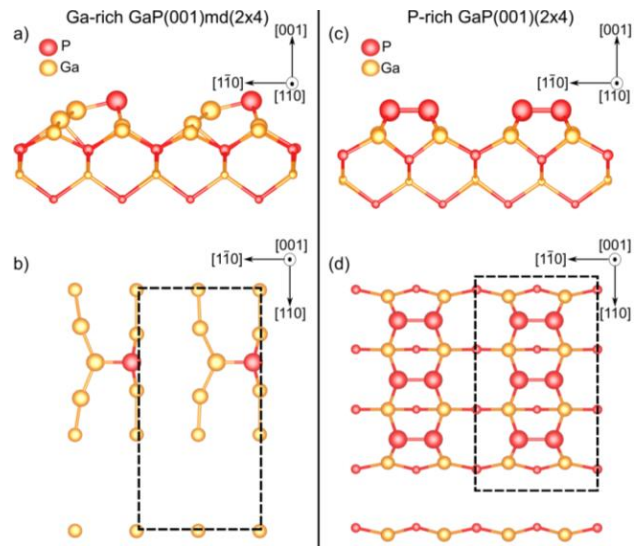

**Fig. S2:** GaP(001) surfaces reconstructions investigated**:** a) side view and b) top view of the Ga-rich GaP(001)md(2x4) surface while c) side view and d) top view of the P-rich GaP(001)(2x4) surface. The surfaces reconstructions unit cells are surrounded by dashed lines in the top views.

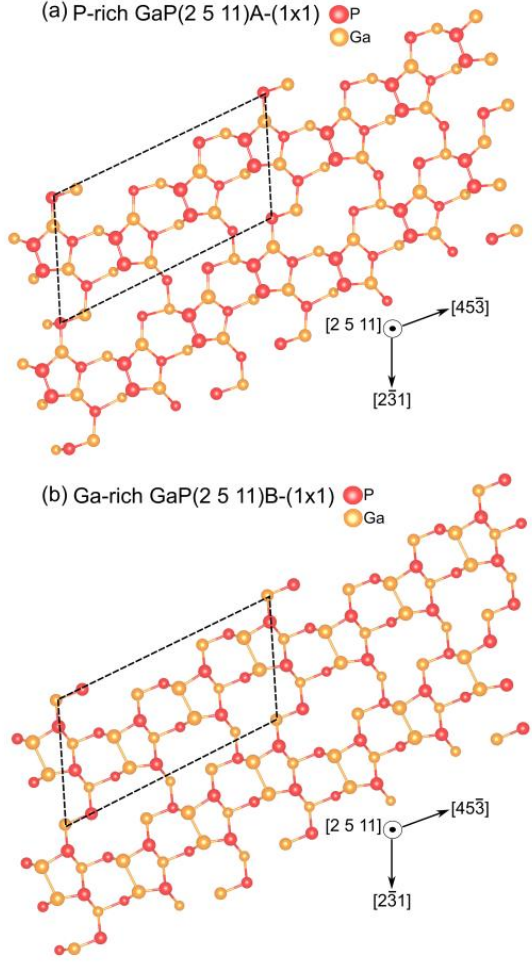

**Fig. S3:** a) P-rich GaP(2 5 11)A-(1x1) b) Ga-rich GaP(2 5 11)B-(1x1) surfaces reconstructions top views. Unit cells in dashed lines.

Surface energies calculations

The non-polar surfaces energies have been calculated without any fictitious H$^*$ atoms on the bottom surface of the slab to provide a more accurate value on the energy. The top and bottom surfaces have been checked to be the same to a rotation when the minimum of energy has converged. In such a case, the relation to calculate the surface energy is:

$$\gamma_{non-polar} = \frac{E_{slab} - N_{Ga}\mu_{GaP} - (N_P - N_{Ga})\mu_P}{2A} \quad (S2)$$

where $E_{slab}$ is the slab energy when the reconstruction is achieved, $N_i$ is the number of particles of the species $i$ (where i is Ga, or P), $\mu_{GaP}$ and $\mu_P$ are the chemical potential of the GaP material and of the species P and $2A$ is the two surface reconstruction areas (top and bottom).

The non-polar GaP(001) surface investigated are non-stoichiometric, i.e., $N_P \neq N_{Ga}$. It means that the surface energy strongly depends on its stoichiometry $\Delta N = N_P - N_{Ga}$ and consequently also on the variation on the chemical potential $\mu_P$. Therefore, it is important to define the chemical potential which should behave in accordance with the thermodynamic conditions.

Indeed, the chemical potentials $\mu_P$ and $\mu_{Ga}$ are defined as the variables that can have each element within the bulk or surface GaP material.

The thermodynamic conditions, to which the chemical potentials have to obey to, are the following: the upper limit of $\mu_P$ and $\mu_{Ga}$ is reached when each element is in its own pure bulk phase:

$$\mu_P < \mu_P^{P-bulk} \quad (S3)$$

$$\mu_{Ga} < \mu_{Ga}^{Ga-bulk} \quad (S4)$$

Moreover, at thermodynamic equilibrium the sum of $\mu_P$ and $\mu_{Ga}$ must be equal to the chemical potential $\mu_{GaP}^{bulk}$ of the GaP bulk phase:

$$\mu_{Ga} + \mu_P = \mu_{GaP}^{GaP-bulk} \quad (S5)$$

$$\mu_{GaP}^{GaP-bulk} = \mu_{Ga}^{Ga-bulk} + \mu_P^{P-bulk} + \Delta H_f(\text{GaP}), \quad (S6)$$

where $\Delta H_f(\text{GaP})$ is the GaP heat of formation. Its value of -0.928 eV has been determined, according to the literature [33–35].
In this work, we therefore express the GaP(001) surface energies as a function of the phosphorus chemical potential variation $\Delta\mu_P = \mu_P - \mu_P^{P-bulk}$. Thus, by combining (S3), (S5) and (S6), the extreme thermodynamic conditions for $\Delta\mu_P$ are given by:

$$\Delta H_f(GaP) < \Delta\mu_P < 0 \quad (S7)$$

To summarize, when $\Delta\mu_P$ equals to the heat formation $\Delta H_f(GaP)$, extreme Ga-rich limit is reached (i.e. bulk Ga phase will form preferentially). When $\Delta\mu_P$ equals 0, the extreme P-rich limit is reached (i.e. bulk P phase will form preferentially).

For the polar surface energy, we first applied on the bottom of the slab the technique of the fictitious H$^*$-passivated surface [36] which has been fruitfully demonstrated on a similar semiconductor GaAs crystal [37,38]. The relation to calculate the surface energy is therefore modified by including two new terms $N_{H^*}^{Ga}\mu_{H^*}^{Ga} + N_{H^*}^{P}\mu_{H^*}^{P}$ coming from the fictitious H$^*$ atoms:

$$\gamma_{polar} = \frac{E_{slab} - N_{Ga}\mu_{GaP} - (N_P - N_{Ga})\mu_P}{A} \\ \frac{-N_{H^*}^{Ga}\mu_{H^*}^{Ga} - N_{H^*}^{P}\mu_{H^*}^{P}}{A} \quad (S8)$$

where $N_{H^*}^i$ and $\mu_{H^*}^i$ are the number and the chemical potential of fictitious H$^*$ related to the species $i$ (where i is Ga, or P), and $A$ is the surface reconstruction area (top). For the GaP(2 5 11) surface, $N_{H^*}^{Ga}$ and $N_{H^*}^{P}$ numbers of fictitious H$^*$ atoms are exactly equals so we can rename it as $N_{H^*}$. Now, the main issue is to evaluate the value of the sum $\mu_{H^*}^{Ga} + \mu_{H^*}^{P}$ of the two chemical potentials of the fictitious H$^*$ atoms. Finally, to be able to approximate the surface energy of the polar surface, slabs of GaP(2 5 11) have been built on the bulk position of GaP material then the dangling bonds on top and bottom surfaces of GaP(2 5 11) were passivated with the appropriate fictitious H$^*$ atom as explained above. Then, the fictitious H$^*$ atoms were only allowed to relax and the Ga and P atoms were kept frozen in the bulk position. The sum of the chemical potentials is approximated by this relation:

$$\mu_{H^*}^{Ga} + \mu_{H^*}^{P} = \frac{E_{slab}^{H^*-passivated} - N_{GaP}\mu_{GaP}}{N_{H^*}} \quad (S9)$$

where $E_{slab}^{H^*-passivated}$ is the energy of the slab when only the fictitious H$^*$ atoms on the top and bottom have been minimized, $N_{GaP}$ is the number of GaP pair within the slab and $N_{H^*}$ and $\mu_{GaP}$ have been already defined above.

To our experience, the sum $\mu_{H^*}^{Ga} + \mu_{H^*}^{P}$ of the two chemical potentials of the fictitious H$^*$ atoms highly depends on the studied surface and should not have the same value from one surface to another.

For convergence reasons and to validate the method, we first calculated the surface energy for the polar GaAs(114) as a test to compare with the previous study [38]. Indeed, we found an identical surface energy for Ga-rich GaAs(114)A-α2(2x1) reconstruction.

In Table S2, the surface energy of non-polar and polar GaP surfaces calculated with the two methods explained above are reported.

**Table S2**. Surface energies of GaP(001) and GaP(2 5 11) computed by DFT for non-polar and polar.

| GaP surface energies γ | P-rich GaP(001) (2x4) | Ga-rich GaP(001) md (2x4) | P-rich GaP (2 5 11) (1x1) | Ga-rich GaP (2 5 11) (1x1) |
|---|---|---|---|---|
| Ga-rich | 72.4 | 52.9 | 62.7 | 57.1 |
| P-rich | 57.4 | 82.8 | 52.9 | 66.8 |

## GAP-SI HETERO-INTERFACE

### GaP/Si interface energies

So far, the GaP/Si interface energy has already been investigated by previous works. Indeed, results on the relative interface formation energy of the GaP on different Si surfaces has been already presented in Ref. [33]. The stability of the compensated GaP/Si(001) interface with respect to an abrupt one has been reported as well in references [34,39] by calculating its relative formation energy. The GaP/Si(001) absolute abrupt interface energy has also been determined [40] but these results have been considered incorrect, as commented in reference [41], since the dependence from the chemical potential of the absolute interface energy has not been considered. So finally, a correct value of the GaP/Si(001) absolute abrupt interface energy has not been found yet.

Our DFT calculations to determine the GaP/Si(001) absolute abrupt interface energy as a function of the chemical potential are presented in the following.

The calculations computed by DFT have been done using the same parameters already reported in the paragraph above for the GaP(001) surfaces energies study.

### GaP/Si slabs

To determine the interface energies, we studied both the abrupt P/Si and Ga/Si interface. The slabs are shown in Fig. S4e-h. For each interface, the top surface is modeled by stable reconstructions studied for the GaP(001): Ga-rich GaP(001)md (2x4) or P-rich GaP(001)(2x4). The unit cell of each one is shown in the top view section in Fig. S4b,d and Fig. S4a,c respectively. As already mentioned above, these surfaces obey to the ECM. Due to different construction arrangements taken into account to build the slabs, the P-rich (Ga-rich) unit cells of Fig. S4a,c (Fig. S4b,d) are not the same. The slabs are separated by a vacuum region of 450Å thick. To avoid any surface/interface interaction, both the GaP and Si bulk are 20Å thick each. More precisely, the slab length in Fig. S4e,f is respectively 42.31Å, 43.62Å while the slabs in Fig. S4g,h have respectively a length of 40.9Å and 45Å. For each slab, the basis vectors length is 15.44Å and 7.72Å. We choose the Si(001) as bottom surface of each case investigated. Finally, the entire GaP together with the two first layers of Silicon at the interface were relaxed, while the rest has been frozen.

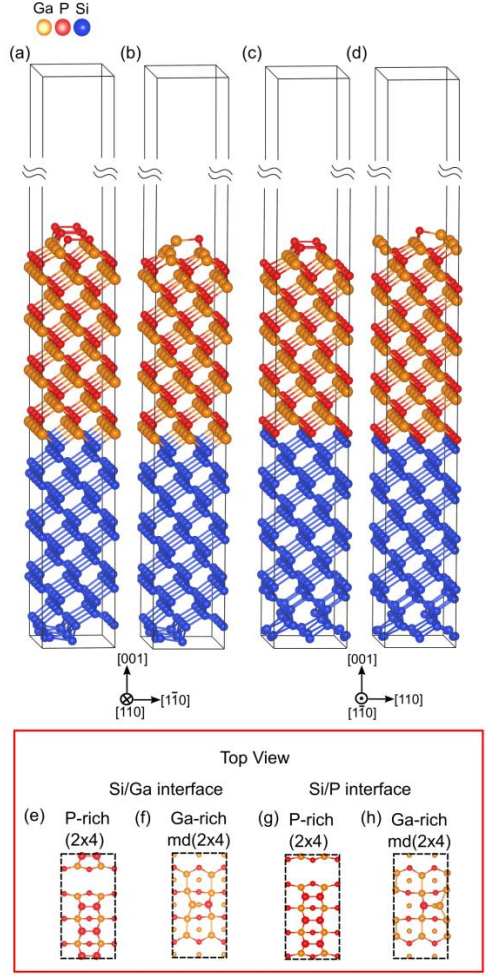

**Fig. S4:** Slabs for DFT calculation for the absolute interface energy and their top view. a) P-rich GaP(001)(2x4) and b) Ga-rich GaP(001)md (2x4) surfaces reconstructions correspond to the Ga-Si interface. The surfaces top views are e) and f) respectively. c) P-rich GaP(001)(2x4) and d) Ga-rich GaP(001)md (2x4) surfaces reconstructions correspond to the P-Si interface and g) and h) are their top view.

Calculations of interface energies

The interface energy $^X\gamma_Z^Y$, in meV/Å², has been defined with X as the study interface. Y and Z are the top and bottom specific surfaces of the slab related to the 2 considered materials. Here, the X interfaces for Silicon and GaP materials are Si-Ga or Si-P. The equation is the following:

$$^X\gamma_Z^Y = \frac{E_{slab}^{int} - \sum_{i=Y,Z}(E_{bulk}^i + A\gamma_{surf}^i)}{A} \quad (S10)$$

Where $E_{slab}^{int}$ is the total energy of the slab, $E_{bulk}^i$ and $\gamma_{surf}^i$ are respectively the energy of the bulk material i and the specific surface energy for material i (with Y and Z). Then, we can rewrite this overall relation to our specific case such as:

$$^X\gamma_Z^Y = \frac{E_{slab}^{int} - N_{Ga}\mu_{GaP} - (N_P - N_{Ga})\mu_P}{A}$$
$$\frac{-N_{Si}E_{bulk}^{Si} - A\gamma_{surf}^{Si} - A\gamma_{surf}^{GaP}}{A} \quad (S11)$$

Where $E_{slab}^{int}$ is defined above, $N_{Ga}$ and $N_P$ are respectively the number of Ga and P atoms of the slab investigated, $\mu_{GaP}$ and $\mu_P$ are the chemical potentials of the GaP and of the species P and A is the rectangular base surface area. $E_{bulk}^{Si}$ is the silicon bulk energy while $N_{Si}$ is the number of silicon atoms. $\gamma_{surf}^{Si}$ and $\gamma_{surf}^{GaP}$ are the specific bottom and the top surface energy per unit area.

The chemical potential of species P varies in the same interval range of the GaP surfaces case.

The results are shown in Table S3. The Si-Ga interface is always more stable in Ga-rich environment while the Si-P interface is stable in the P-rich one. Moreover, this is independent of the kind of surface considered with small numerical error except for Si-P where this error is increasing. However, the absolute variation of the interface energy from P-rich to Ga-rich conditions is always of 31.2meV/Å² for Si-Ga and Si-P interface.

**Table S3** GaP/Si interface energies computed by DFT

| Interface | Energy (meV/Å²) | |
|---|---|---|
|  | P-rich | Ga-rich |
| $^{Si-Ga}\gamma_{GaP\,(001)\,2x4}^{Si\,(100)}$ | 72.0 | 40.8 |
| $^{Si-Ga}\gamma_{GaP\,(001)\,2x4md}^{Si\,(100)}$ | 69.7 | 38.5 |
| $^{Si-P}\gamma_{GaP\,(001)\,2x4}^{Si\,(100)}$ | 29.7 | 60.9 |
| $^{Si-P}\gamma_{GaP\,(001)\,2x4md}^{Si\,(100)}$ | 23.3 | 54.5 |

# FREE ENERGY CALCULATIONS

The total free energy variation during the GaP/Si growth is calculated here. It corresponds to the difference of free energy between an initial thermodynamic state with a total atom N related to the sum of Ga and P atoms in a vapor reservoir and with a Si substrate, and a final state where the GaP crystal is formed onto the Si. (see ref. [42] for a precise description of the process.) Free energy variations were then calculated for the different 2D or 3D islands configurations by using [42]:

$$\Delta F_{TOT} = \Delta F_\mu + \Delta F_e + \Delta F_{S\&i} \quad (S12)$$

The first term is the chemical work needed to form the bulk crystal from an infinite reservoir. For Molecular Beam Epitaxy (MBE) of GaP using a $P_2$ source, it comes:

$$\Delta F_\mu = Nk_B T \ln\left(\frac{P_{Ga}\cdot(P_{P_2})^{1/2}}{P_{Ga-\infty}\cdot(P_{P_2-\infty})^{1/2}}\right) \quad (S13)$$

Where T is the growth temperature, N the number of condensed atoms, $P_X$ the partial pressure of species X, $P_{X-\infty}$ the saturation partial pressure of species X, and $k_B$ the Boltzmann constant. While T and $P_X$ are extracted directly from growth conditions, the saturation pressures have been precisely calibrated in ref. [43], section 2.5.4 for GaP.

The second term is associated to the elastic energy stored and is defined as:

$$\Delta F_e = \mathcal{F}_0 m^2 V R \quad (S14)$$

Where m is the epitaxial misfit between the deposited material and the substrate, V the volume of the deposited crystal, $\mathcal{F}_0$ a combination of the elastic coefficients $C_{ij}$ and R a relaxation energy factor, that traduces the strain status of the deposited crystal and the substrate [42]. R=0 corresponds to a perfectly relaxed system, while R=1 correspond to a perfectly non-relaxed crystal. In the present work, for GaP/Si, we consider either $\Delta F_e$-2D a perfect biaxial strain along the GaP epilayers (R=1) that is typically expected for a 2D GaP island growing on Si, or a relaxation due to 3D islanding $\Delta F_e$-3D, for which we consider R=0.7 [44].

We note here that R depends on the island shape. The energy gain provided by the transition of an equilibrium Wulff-Kaishew island (R=0.7) and a similar non-truncated island (R=0.6) is not sufficient to compensate the corresponding surface energy increasing. This also applies for an island with {111} facets, where R=0.3. Therefore, the gain provided by elastic relaxation is always several orders of magnitude lower than the corresponding surface/interface energy cost and therefore won't have any influence on the island shape.

Finally, for a cubic crystal stressed in a (001) plane, $\mathcal{F}_0$ is expressed as: $(C_{11} + C_{12} - 2\frac{C_{12}^2}{C_{11}})$.

The third term corresponds to the formation of surfaces and interfaces, which rewrites in the present case:

$$\Delta F_{S\&i} = \begin{array}{l} \sum_j \gamma^S_{(III-V),j} \cdot S_{(III-V),j} \\ +S_{(III-V/Si)}(\gamma^i_{(III-V/Si)} - \gamma^S_{(Si)}) \end{array} \quad (S15)$$

Where $\gamma^S_{(III-V),j}$ and $\gamma^S_{(Si)}$ are the surface energies of the j[th] III-V facet and of the silicon surface respectively, $\gamma^i_{(III-V/Si)}$ is the interface energy between the III-V semiconductor and the Si, $S_{(III-V),j}$ the surface of the j[th] III-V facet and $S_{(III-V/Si)}$ the contact surface between the III-V and the Si. In this work, we neglect the vibrational contribution to the free energy which is not expected to impact the main conclusions [45].

# REALISTIC GEOMETRY OF GAP ISLANDS

A careful STM image analysis has been performed on the data of Fig. 1(b) in the paper, which gives an average island height of 2.5 nm, and an average diameter of 11 nm, that leads to an average (miscut included) island contact angle of 27.04°. Among the different stable facets observed with GaP or GaAs materials that are mainly lying around the {001}, {111}, {136} and {114} ones [5,46], the measured contact angle only corresponds to {136} ones (theoretical contact angle of 27.8°).

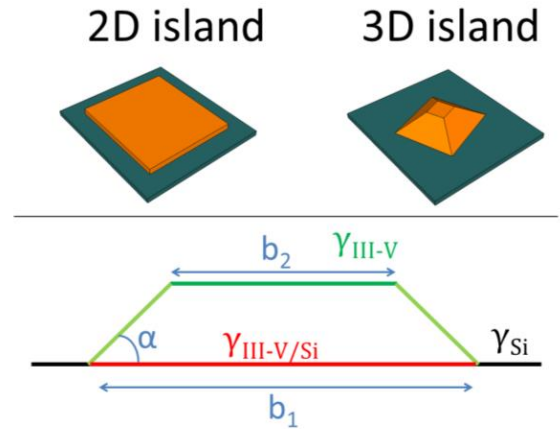

**Fig. S5:** Sketch of the 2D (strained) and 3D (elastically relaxed) GaP islands on Si. The parameters (b1, b2 and α) defining the shape of the truncated pyramids are represented.

We therefore model the GaP 3D islands by truncated pyramidal structures composed of facets with an angle

$\alpha=27.8°$, having the surface energy of {136} facets. As explained in previous sections, the surface energy of this facet is taken from the {2511} one as it is the most stable configuration, and respect the electron counting model.

As described in Fig. S5, the pyramid has a square basis, a {001} facet on top, and grows in a homothetic way during the initiation steps. Truncated pyramid islands are chosen at their equilibrium shape determined by the Wulf–Kaishew theorem. [42] In P-rich conditions, $b_2/b_1=0.05$, while in Ga-rich conditions, $b_2/b_1=0.6$. For the modeling of the 2D GaP island on Si, we model the top surface by a conventional {001} facet and keep a one monolayer height thickness; the 2D island is only growing laterally. The edge energy is neglected, giving a lower limit estimated around $10^3$ for the total number of atoms composing the island.